Letter

# Revealing dipion correlations for the observed substructure near the $\pi^+\pi^-$ mass threshold in $\psi(3686) \to J/\psi\pi^+\pi^-$

Zhong-Yu Wang,[1,*] Zhe Liu,[2,3,4,5,†] and Xiang Liu[2,3,4,6,5,‡]

[1]College of Physics, Guizhou University, Guiyang 550025, China
[2]School of Physical Science and Technology, Lanzhou University, Lanzhou 730000, China
[3]Lanzhou Center for Theoretical Physics, Key Laboratory of Theoretical Physics of Gansu Province, Lanzhou University, Lanzhou 730000, China
[4]Key Laboratory of Quantum Theory and Applications of MoE, Lanzhou University, Lanzhou 730000, China
[5]Research Center for Hadron and CSR Physics, Lanzhou University and Institute of Modern Physics of CAS, Lanzhou 730000, China
[6]MoE Frontiers Science Center for Rare Isotopes, Lanzhou University, Lanzhou 730000, China

(Received 13 November 2025; accepted 22 January 2026; published 12 February 2026)

Based on the world's largest $\psi(3686)$ data sample, the BESIII Collaboration recently reported a substructure near the $\pi^+\pi^-$ mass threshold in the decay $\psi(3686) \to J/\psi\pi^+\pi^-$, challenging the established understanding of the dipion invariant mass spectrum. We propose that this substructure arises directly from dipion correlations. Using a chiral unitary approach, we successfully reproduce the observed anomaly, thereby providing strong evidence of dipion correlation in heavy quarkonium decays. This approach also allows us to predict the corresponding dipion correlation function.

Matter correlations are an interesting and universal phenomenon, spanning from the microscopic to the macroscopic scale. They are a direct result of interactions, and thus, investigating them can reveal how matter interacts. Among these, hadron correlations have received significant attention in recent years [1–16], driven by the accumulation of experimental data. A primary objective of this research is to deepen our understanding of the nonperturbative aspects of the strong interaction, which is a central problem in modern particle physics.

Very recently, the BESIII Collaboration reported a substructure near the $\pi^+\pi^-$ mass threshold in the decay $\psi(3686) \to J/\psi\pi^+\pi^-$ [17]. Earlier measurements of the dipion invariant mass spectrum in this decay [18–20] have motivated the application of the QCD multipole expansion (QCDME) method, as proposed in Ref. [21], to study low-lying heavy quarkonium transitions. The process $\psi(3686) \to J/\psi\pi^+\pi^-$ represents a case where the QCDME approach worked particularly well [22]. In recent years, the BESIII detector accumulated the world's largest $\psi(3686)$ data sample, comprising $(2712.4 \pm 1.4) \times 10^6$ events, which enabled the first observation of a substructure near the $\pi^+\pi^-$ mass threshold in this decay channel [17,23]. This finding highlights the potential of high-precision hadron spectroscopy.

This unexpected observation challenges the established understanding of the dipion invariant mass spectrum in $\psi(3686) \to J/\psi\pi^+\pi^-$. Building on these experimental results, Ref. [24] investigated this decay process using dispersion theory and demonstrated that no additional resonance state is required. We propose that the observed substructure near the $\pi^+\pi^-$ mass threshold originates directly from dipion correlations. Using a chiral unitary approach [25–29], we successfully reproduce the substructure, providing strong evidence for dipion correlations in heavy quarkonium decays.

We present the theoretical formalism for the reaction $\psi(3686) \to J/\psi\pi^+\pi^-$ via final-state interactions. Given that both the initial state $\psi(3686)$ and the final state $J/\psi$ have spin $J = 1$, while the $\pi^+$ and $\pi^-$ in the final state have spin $J = 0$, we construct the following decay amplitude structure

$$t = \epsilon_\mu(\psi(3686))\epsilon^{*\mu}(J/\psi)\tilde{t}_{\pi^+\pi^-}, \tag{1}$$

where $\tilde{t}_{\pi^+\pi^-}$ denotes the $S$-wave transition amplitude for $\pi^+\pi^-$ scattering within coupled channels. In principle, an $S$-wave interaction exists between $J/\psi$ and $\pi^+$ (or $\pi^-$), but its contribution to the low-energy region of the $\pi^+\pi^-$ invariant mass distribution is negligible. Moreover, interactions between $J/\psi$ and $\pi^+(\pi^-)$ are suppressed by the Okubo-Zweig-Iizuka (OZI) rule and are therefore disregarded.

*Contact author: zhongyuwang@gzu.edu.cn
†Contact author: zhliu20@lzu.edu.cn
‡Contact author: xiangliu@lzu.edu.cn

To identify the appropriate combination of three mesons in the $\psi(3686)$ decay, we introduce the quark-level flavor-space matrix $M = (q_i\bar{q}_j)$, i.e.,

$$M = \begin{pmatrix} u\bar{u} & u\bar{d} & u\bar{s} & u\bar{c} \\ d\bar{u} & d\bar{d} & d\bar{s} & d\bar{c} \\ s\bar{u} & s\bar{d} & s\bar{s} & s\bar{c} \\ c\bar{u} & c\bar{d} & c\bar{s} & c\bar{c} \end{pmatrix}. \tag{2}$$

The properties of this matrix are detailed in Refs. [30,31]. For the initial $\psi(3686)$ state, which is composed of a $c\bar{c}$ pair and represented by the (4,4) element of $M$, the physically accessible final-state configurations arise from terms such as $(M\cdot M\cdot M)_{44}$, $(M\cdot M)_{44}\mathrm{Tr}(M)$, $(M)_{44}\mathrm{Tr}(M\cdot M)$, and $(M)_{44}\mathrm{Tr}(M)\mathrm{Tr}(M)$. Specifically, for the decay $\psi(3686)\to J/\psi\pi^+\pi^-$, where final-state interactions are confined to the $\pi\pi$ subsystem and $J/\psi\pi$ interactions are absent, only the last two OZI-suppressed terms—$(M)_{44}\mathrm{Tr}(M\cdot M)$ and $(M)_{44}\mathrm{Tr}(M)\mathrm{Tr}(M)$—contribute. This contribution arises from replacing the quark-level matrix $M$ in these terms with the pseudoscalar meson matrix $P$ and the vector meson matrix $V^\mu$, yielding $(V^\mu)_{44}\mathrm{Tr}(P\cdot P)$ and $(V^\mu)_{44}\mathrm{Tr}(P)\mathrm{Tr}(P)$. The matrices $P$ and $V^\mu$ are defined as,

$$P = \begin{pmatrix} \frac{\eta}{\sqrt{3}}+\frac{\eta'}{\sqrt{6}}+\frac{\pi^0}{\sqrt{2}} & \pi^+ & K^+ & \bar{D}^0 \\ \pi^- & \frac{\eta}{\sqrt{3}}+\frac{\eta'}{\sqrt{6}}-\frac{\pi^0}{\sqrt{2}} & K^0 & D^- \\ K^- & \bar{K}^0 & -\frac{\eta}{\sqrt{3}}+\sqrt{\frac{2}{3}}\eta' & D_s^- \\ D^0 & D^+ & D_s^+ & \eta_c \end{pmatrix}, \tag{3}$$

$$V^\mu = \begin{pmatrix} \frac{\omega+\rho^0}{\sqrt{2}} & \rho^+ & K^{*+} & \bar{D}^{*0} \\ \rho^- & \frac{\omega-\rho^0}{\sqrt{2}} & K^{*0} & D^{*-} \\ K^{*-} & \bar{K}^{*0} & \phi & D_s^{*-} \\ D^{*0} & D^{*+} & D_s^{*+} & J/\psi \end{pmatrix}^\mu. \tag{4}$$

The term $(M)_{44}\mathrm{Tr}(M\cdot M)$ contributes to the process depicted in Fig. 1(a), yielding the following three-meson combination

$$|H^{(1)}\rangle = 2J/\psi\pi^+\pi^- + J/\psi\pi^0\pi^0 + 2J/\psi K^+K^- + 2J/\psi K^0\bar{K}^0 + J/\psi\eta\eta. \tag{5}$$

The term $(M)_{44}\mathrm{Tr}(M)\mathrm{Tr}(M)$ contributes to the process in Fig. 1(b), giving

$$|H^{(2)}\rangle = \frac{1}{3}J/\psi\eta\eta. \tag{6}$$

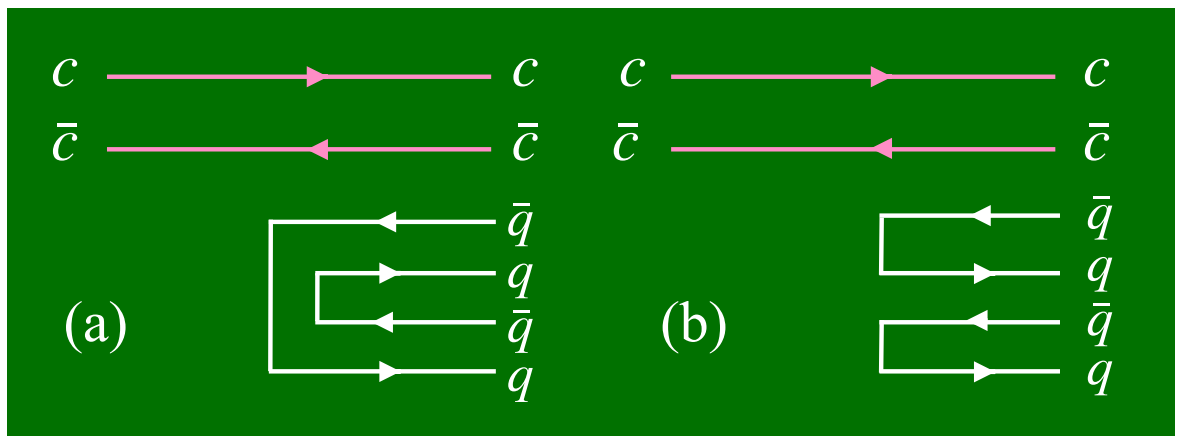

FIG. 1. Topological diagrams corresponding to the (a) $(M)_{44}\mathrm{Tr}(M\cdot M)$ and (b) $(M)_{44}\mathrm{Tr}(M)\mathrm{Tr}(M)$ terms.

For Eq. (5), the decay $\psi(3686)\to J/\psi\pi^+\pi^-$ proceeds through both tree-level and $S$-wave pseudoscalar-pseudoscalar interactions, as shown in Fig. 2(a). For Eq. (6), the $\pi^+\pi^-$ pair in the final state originates solely from the $\eta\eta$ channel via final-state interactions during the initial production stage. Figure 2(b) illustrates the corresponding rescattering contributions.

The total amplitude is expressed as

$$\begin{aligned} \tilde{t}_{\pi^+\pi^-}(M_{23}) = V_1[&2 + 2G_{\pi^+\pi^-}(M_{23})T_{\pi^+\pi^-\to\pi^+\pi^-}(M_{23}) \\ &+ \sqrt{2}G_{\pi^0\pi^0}(M_{23})T_{\pi^0\pi^0\to\pi^+\pi^-}(M_{23}) \\ &+ 2G_{K^+K^-}(M_{23})T_{K^+K^-\to\pi^+\pi^-}(M_{23}) \\ &+ 2G_{K^0\bar{K}^0}(M_{23})T_{K^0\bar{K}^0\to\pi^+\pi^-}(M_{23}) \\ &+ \sqrt{2}G_{\eta\eta}(M_{23})T_{\eta\eta\to\pi^+\pi^-}(M_{23})] \\ &+ \frac{\sqrt{2}}{3}V_2e^{i\phi}G_{\eta\eta}(M_{23})T_{\eta\eta\to\pi^+\pi^-}(M_{23}), \end{aligned} \tag{7}$$

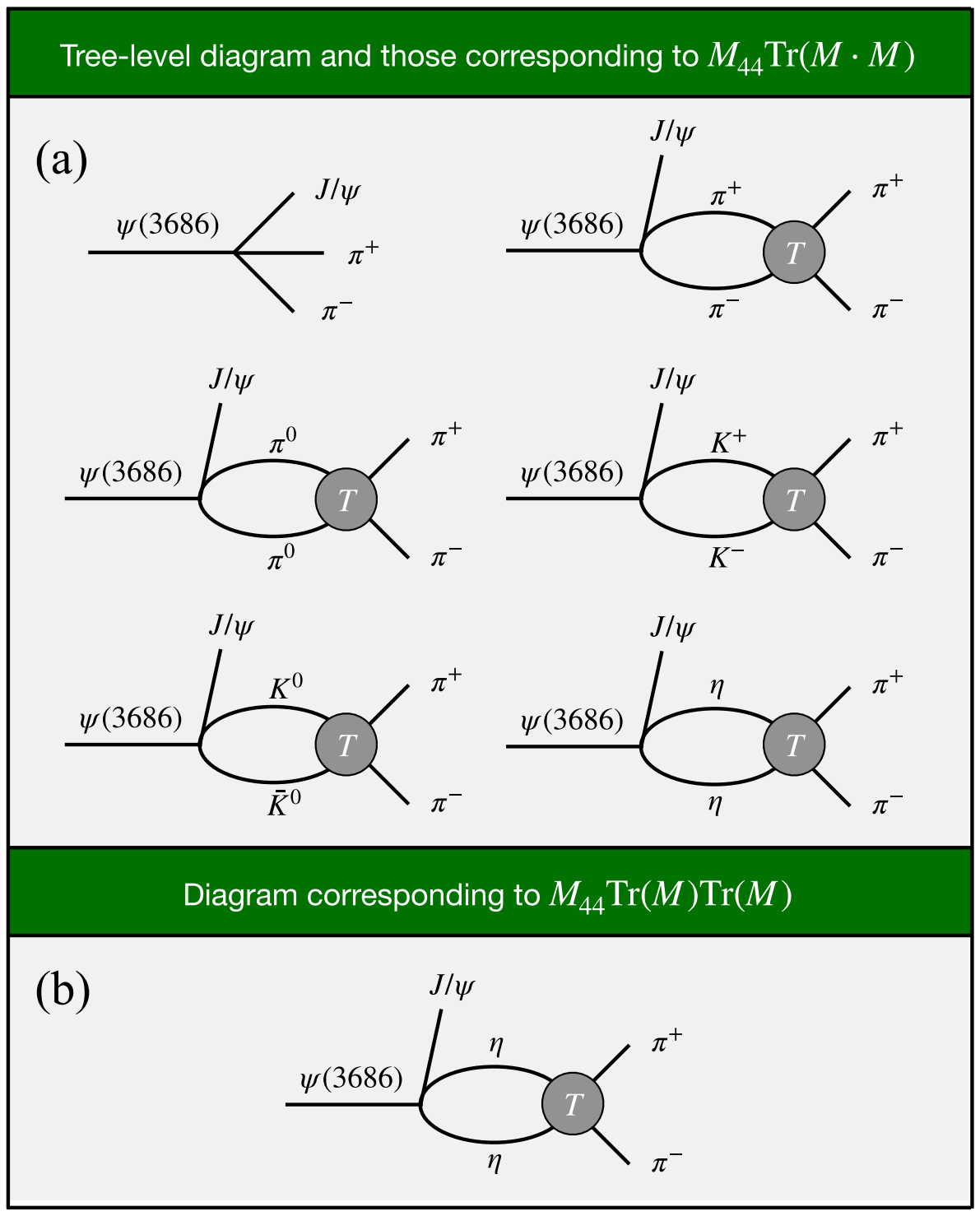

FIG. 2. Diagrams contributing to $\psi(3686)\to J/\psi\pi^+\pi^-$.

where $V_1$ and $V_2$ are the decay strengths from Eqs. (5) and (6), respectively. A relative phase $\phi$ is introduced between these two decay mechanisms. We treat these parameters as constants and determine their values by fitting to the BESIII experimental data. Here, $M_{ij}$ denotes the invariant mass of a meson pair, with subscripts 12, 13, and 23 corresponding to the final-state particle pairs $J/\psi\pi^+$, $J/\psi\pi^-$, and $\pi^+\pi^-$, respectively. A factor of 2 arises in identical-particle channels $\pi^0\pi^0$ and $\eta\eta$, cancelling the $1/2$ factor in their loop function. The additional factor of $\sqrt{2}$ in these channels stems from the unitary normalization used in the chiral unitary approach: $|\pi^0\pi^0\rangle \to \frac{1}{\sqrt{2}}|\pi^0\pi^0\rangle$ and $|\eta\eta\rangle \to \frac{1}{\sqrt{2}}|\eta\eta\rangle$. For further discussion, see Refs. [25,32].

The loop function $G$ for two mesons is expressed using the three-momentum cutoff formulation from Ref. [25]

$$G_{ii}(s) = \int_0^{q_{\max}} \frac{q^2 dq}{(2\pi)^2} \frac{\omega_1+\omega_2}{\omega_1\omega_2[s-(\omega_1+\omega_2)^2+i\epsilon]}, \quad (8)$$

where $\omega_i(\vec{q}) = \sqrt{\vec{q}^{\,2}+m_i^2}$, with $m_1$ and $m_2$ being the masses of the two mesons in the coupled channels, and $q_{\max}$ is a free parameter. The loop function in dimensional regularization is given by [29,33–36]

$$\begin{aligned} G_{ii}(s) = \frac{1}{16\pi^2}\Bigg\{ & a_i(\mu) + \ln\frac{m_1^2}{\mu^2} + \frac{m_2^2-m_1^2+s}{2s}\ln\frac{m_2^2}{m_1^2} \\ & + \frac{q_{cmi}(s)}{\sqrt{s}}[\ln(s-(m_2^2-m_1^2)+2q_{cmi}(s)\sqrt{s}) \\ & + \ln(s+(m_2^2-m_1^2)+2q_{cmi}(s)\sqrt{s}) \\ & - \ln(-s-(m_2^2-m_1^2)+2q_{cmi}(s)\sqrt{s}) \\ & - \ln(-s+(m_2^2-m_1^2)+2q_{cmi}(s)\sqrt{s})]\Bigg\}, \end{aligned} \quad (9)$$

where $\mu$ is the renormalization scale, $a_i(\mu)$ is the subtraction constant, and $q_{\rm cmi}(s)$ is the center-of-mass momentum

$$q_{cmi}(s) = \frac{\lambda^{1/2}(s, m_1^2, m_2^2)}{2\sqrt{s}} \quad (10)$$

with $\lambda(a,b,c) = a^2+b^2+c^2-2(ab+ac+bc)$.

Due to singularities in the three-momentum cutoff scheme above threshold, we adopt dimensional regularization. To minimize free parameters, the subtraction constant $a_i(\mu)$ is determined by matching the loop functions of both schemes at the threshold of a given channel, as in Refs. [37–41]

$$a_i(\mu) = 16\pi^2[G^{CO}(s_{thr}, q_{\max}) - G^{DR}(s_{thr}, \mu)], \quad (11)$$

where $G^{\rm CO}$ and $G^{\rm DR}$ are defined by Eqs. (8) and (9), respectively. We set $q_{\max} = \mu = 0.6$ GeV, determined by fitting to experimental data for the $f_0(500)$ and $f_0(980)$ states [30,42]; this value is widely used in subsequent theoretical works.

We consider coupled-channel interactions for isospin $I=0$, including five channels: $\pi^+\pi^-$, $\pi^0\pi^0$, $K^+K^-$, $K^0\bar{K}^0$, and $\eta\eta$. The transition amplitude $T_{i\to\pi^+\pi^-}$ in Eq. (7) is obtained by solving the on-shell Bethe-Salpeter equation [25–27,43]

$$T = [1-VG]^{-1}V, \quad (12)$$

where $V$ represents the $S$-wave interaction for $\pi^+\pi^-$ with its coupled channels, derived from the chiral effective Lagrangian. Explicit forms are given in Refs. [30,44],

$$\begin{aligned} & V_{11} = -\frac{1}{2f^2}s, \quad V_{12} = -\frac{1}{\sqrt{2}f^2}(s-m_\pi^2), \quad V_{13} = -\frac{1}{4f^2}s, \\ & V_{14} = -\frac{1}{4f^2}s, \quad V_{15} = -\frac{1}{3\sqrt{2}f^2}m_\pi^2, \quad V_{22} = -\frac{1}{2f^2}m_\pi^2, \\ & V_{23} = -\frac{1}{4\sqrt{2}f^2}s, \quad V_{24} = -\frac{1}{4\sqrt{2}f^2}s, \quad V_{25} = -\frac{1}{6f^2}m_\pi^2, \\ & V_{33} = -\frac{1}{2f^2}s, \quad V_{34} = -\frac{1}{4f^2}s, \\ & V_{35} = -\frac{1}{12\sqrt{2}f^2}(9s-6m_\eta^2-2m_\pi^2), \quad V_{44} = -\frac{1}{2f^2}s, \\ & V_{45} = -\frac{1}{12\sqrt{2}f^2}(9s-6m_\eta^2-2m_\pi^2), \\ & V_{55} = -\frac{1}{18f^2}(16m_K^2-7m_\pi^2), \end{aligned} \quad (13)$$

where $f = 0.093$ GeV is the pion decay constant.

Finally, the double differential width for the three-body decay is given by [45]

$$\frac{d^2\Gamma}{dM_{12}dM_{23}} = \frac{1}{(2\pi)^3}\frac{1}{8m_{\psi(3686)}^3}M_{12}M_{23}\overline{\sum}\sum|t|^2, \quad (14)$$

and $d\Gamma/dM_{12}$ and $d\Gamma/dM_{23}$ are obtained by integrating over the other invariant mass variable. The distribution $d\Gamma/dM_{13}$ can be derived using the constraint

$$M_{12}^2 + M_{13}^2 + M_{23}^2 = m_{\psi(3686)}^2 + m_{J/\psi}^2 + m_{\pi^+}^2 + m_{\pi^-}^2. \quad (15)$$

We begin by performing a fit to the BESIII experimental data. A new structure located near the $\pi^+\pi^-$ threshold has been reported in the $\pi^+\pi^-$ invariant mass distribution of the decay $\psi(3686) \to J/\psi\pi^+\pi^-$ [17]. Accordingly, we fit the data in the low-energy region of the $\pi^+\pi^-$ invariant mass distribution, specifically in the range of 0.28–0.32 GeV.

Our theoretical model contains three free parameters: $V_1$ and $V_2$, which denote the strengths of the decay mechanisms shown in Fig. 1, respectively (incorporating an overall normalization factor to match the data), and a relative phase $\phi$ between them. The fitted values of these parameters are listed in Table I, yielding $\chi^2/\mathrm{d.o.f.} = 42.86/(45-3) = 1.02$ (d.o.f denotes degrees of freedom).[1]

TABLE I. Theoretical parameters obtained from the fit in the $\psi(3686)$ decay.

| Parameters | $V_1$ ($\times 10^4$) | $V_2$ ($\times 10^9$) | $\phi$ (rad) |
|---|---|---|---|
| Fit values | $4.23 \pm 0.04$ | $1.09 \pm 0.01$ | $0.048 \pm 0.001$ |

Using the fitted parameters, we compute the $\pi^+\pi^-$ invariant mass distribution, as shown in Fig. 3(a). The black points with error bars represent the BESIII experimental data, and the red solid line corresponds to our fit results. Excellent agreement is observed between theory and experiment. A clear peak is seen around 0.285 GeV in the $\pi^+\pi^-$ spectrum, which drops to a minimum near 0.305 GeV. In our model, this structure does not correspond to a genuine resonance, but results from the interference and superposition of the two decay mechanisms illustrated in Fig. 1. The individual contributions of these two components are displayed in Fig. 3(b). Neither shows any structure in this energy region, and both are of comparable magnitude. Above 0.32 GeV, both the data and the theoretical curve rise steadily, dynamically generated from meson-meson interactions. Moreover, we show the $J/\psi\pi^+$ invariant mass distribution in Fig. 3(c). No resonant structures are observed. The $J/\psi\pi^-$ distribution is identical due to charge symmetry.

We further show the full $\pi^+\pi^-$ invariant mass spectrum in Fig. 4. Here, no additional fit was performed; we simply applied the parameters from Table I, adjusting only an overall normalization constant to account for the different event statistics between Figs. 3(a) and 4. This constant affects only the absolute scale, not the spectral shape. The theoretical result agrees well with the experimental data over the full range, with only minor deviations localized around 0.6 GeV. It should be noted that the BESIII data in this region are provided without uncertainties, and the data density is significantly higher than in the low-energy region. For these reasons, we deliberately refrain from fitting the data in Fig. 4. The successful description of the full spectrum using parameters determined from a local fit to the low-energy region demonstrates the robustness and predictive power of our model.

In addition, we calculate the $\pi\pi$ scattering phase shift using $q_{\max} = \mu = 0.6$ GeV, as shown in Fig. 5. Our results agree well with experimental data, exhibiting only minor deviations around 0.95 GeV. This agreement is remarkable given that the phase shift was not fitted directly.

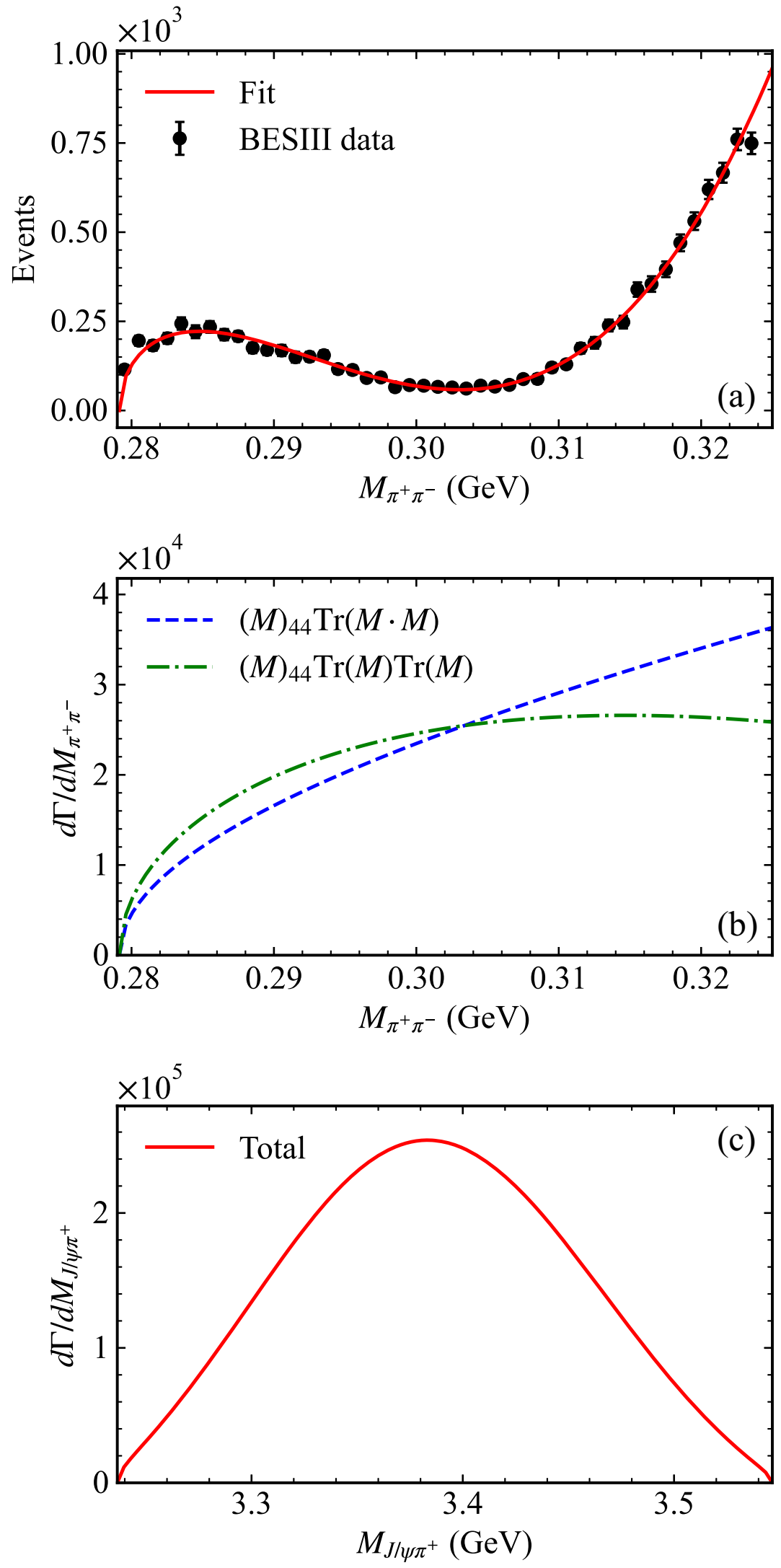

FIG. 3. (a) $\pi^+\pi^-$ invariant mass distribution near threshold in $\psi(3686) \to J/\psi\pi^+\pi^-$ compared with experimental data [17]. (b) Contributions from the two mechanisms in Fig. 1. (c) $J/\psi\pi^+$ invariant mass spectrum.

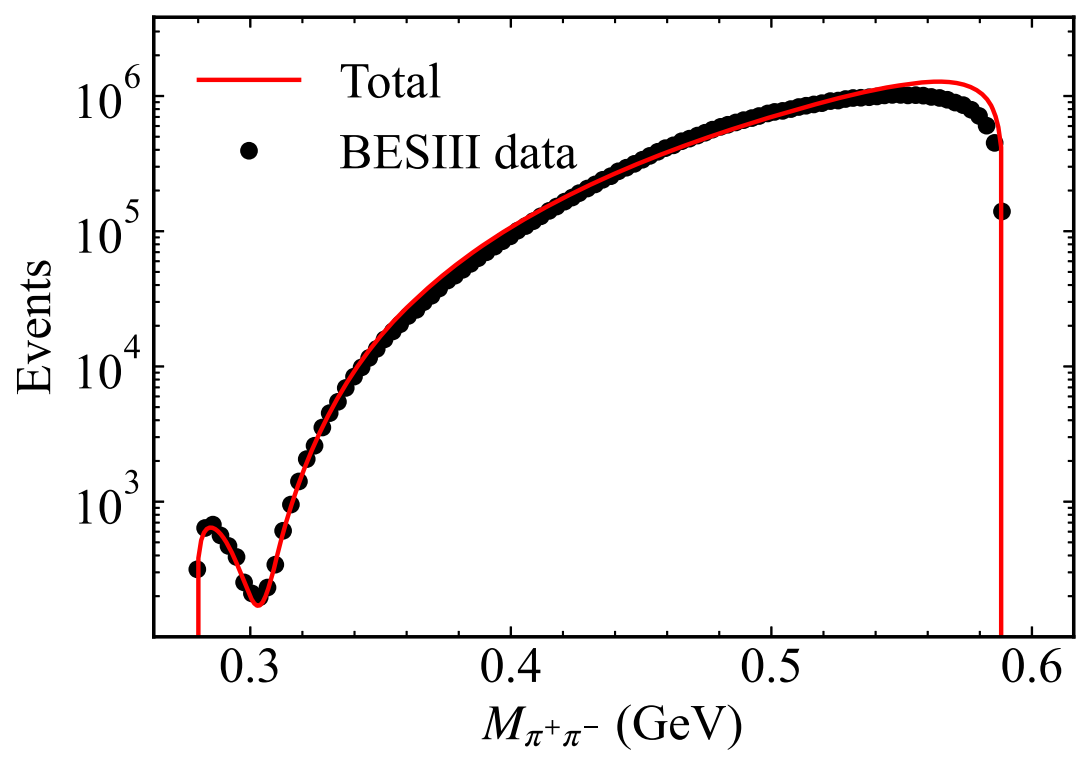

FIG. 4. Full $\pi^+\pi^-$ invariant mass distribution in $\psi(3686) \to J/\psi\pi^+\pi^-$. Data points are from BESIII [17].

[1] In Ref. [17], BESIII investigated the threshold enhancement in the $\pi^+\pi^-$ invariant mass spectrum under a nonresonant hypothesis. Their fit succeeded in reproducing the enhancement; nevertheless, it overshot the spectrum near 0.55 GeV$/c^2$, even as the $\chi^2/\mathrm{d.o.f.}$ was reduced from 310.5 to 16.1. Our work, in contrast, specifically targets the experimental data in the $M_{\pi^+\pi^-} = 0.28$–$0.32$ GeV$/c^2$ region.

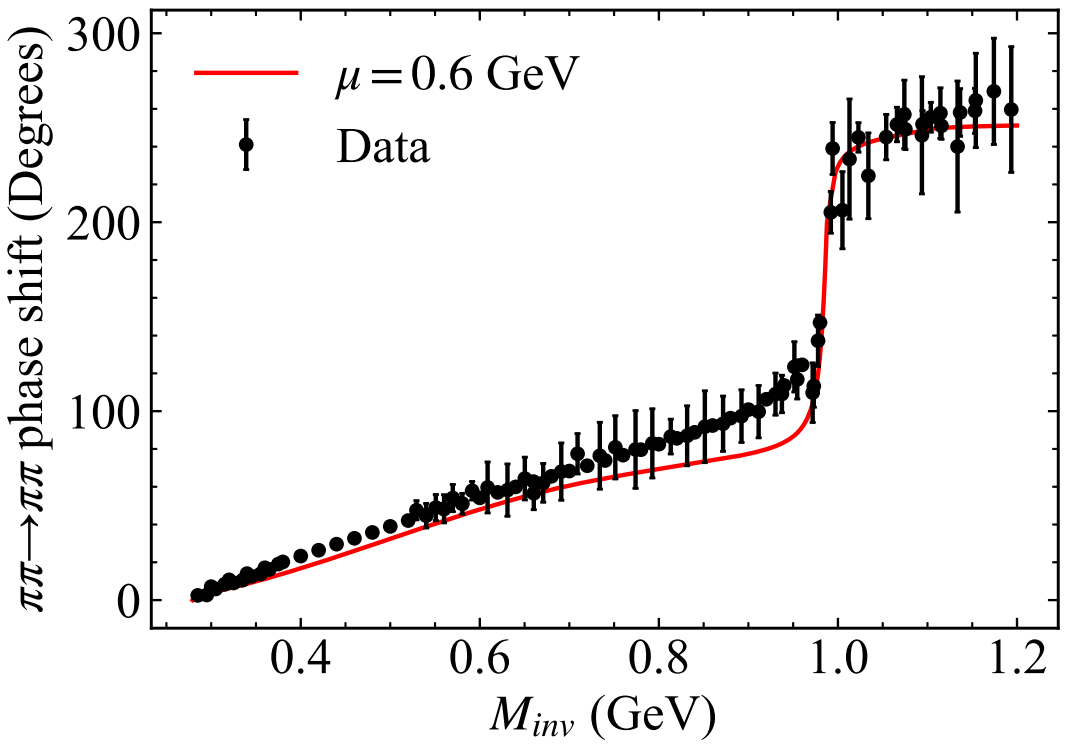

FIG. 5. $I(J) = 0(0)$ $\pi\pi \to \pi\pi$ scattering phase shift obtained with the parameters used in this work. The red curve is our result with $q_{\rm max} = \mu = 0.6$ GeV, and the data without errors are taken from Refs. [46,47], while the ones with errors are the averaged data taken from Refs. [48–50], as used in Refs. [33,51].

Within this framework, the dipion correlation function can also be derived. The coupled-channel correlation functions for both charged and neutral pion pairs are expressed as [52–55]

$$\begin{aligned} C(p_{\pi^+}) = 1 + 4\pi\theta(q_{\rm max} - p_{\pi^+}) \int_0^\infty drr^2 S_{12}(r)[|j_0(p_{\pi^+}r) \\ + T_{\pi^+\pi^-\to\pi^+\pi^-}(s)\tilde{G}_{\pi^+\pi^-}(r,s)|^2 \\ + |\sqrt{2}T_{\pi^0\pi^0\to\pi^+\pi^-}(s)\tilde{G}_{\pi^0\pi^0}(r,s)|^2 \\ + |T_{K^+K^-\to\pi^+\pi^-}(s)\tilde{G}_{K^+K^-}(r,s)|^2 \\ + |T_{K^0\bar{K}^0\to\pi^+\pi^-}(s)\tilde{G}_{K^0\bar{K}^0}(r,s)|^2 \\ + |\sqrt{2}T_{\eta\eta\to\pi^+\pi^-}(s)\tilde{G}_{\eta\eta}(r,s)|^2 - j_0^2(p_{\pi^+}r)], \end{aligned} \quad (16)$$

and

$$\begin{aligned} C(p_{\pi^0}) = 1 + 4\pi\theta(q_{\rm max} - p_{\pi^0}) \int_0^\infty drr^2 S_{12}(r)[|j_0(p_{\pi^0}r) \\ + \sqrt{2}T_{\pi^0\pi^0\to\pi^0\pi^0}(s)\tilde{G}_{\pi^0\pi^0}(r,s)|^2 \\ + |T_{\pi^+\pi^-\to\pi^0\pi^0}(s)\tilde{G}_{\pi^+\pi^-}(r,s)|^2 \\ + |T_{K^+K^-\to\pi^0\pi^0}(s)\tilde{G}_{K^+K^-}(r,s)|^2 \\ + |T_{K^0\bar{K}^0\to\pi^0\pi^0}(s)\tilde{G}_{K^0\bar{K}^0}(r,s)|^2 \\ + |\sqrt{2}T_{\eta\eta\to\pi^0\pi^0}(s)\tilde{G}_{\eta\eta}(r,s)|^2 - j_0^2(p_{\pi^0}r)], \end{aligned} \quad (17)$$

where the factor of $\sqrt{2}$ necessarily originates from the identical particle symmetrization, as formalized in Eq. (7). The parameter $q_{\rm max}$ is still set to 0.6 GeV, as done in Eq. (11). Here, the source function is given by

$$S_{12}(r) = \frac{1}{(R\sqrt{4\pi})^3} e^{-\frac{r^2}{4R^2}}, \quad (18)$$

where $R$ represents the size of the source in which the particles are assumed to be produced. Additionally,

$$\tilde{G}_{ii}(r,s) = \int \frac{d^3\vec{q}}{(2\pi)^3} \frac{\omega_1 + \omega_2}{2\omega_1\omega_2} \frac{j_0(\vec{q}r)}{s - (\omega_1 + \omega_2)^2 + i\epsilon}, \quad (19)$$

where $\omega_i(\vec{q})$ defined as in Eq. (8). The scattering matrix elements $T_{i\to j}$ can be obtained from Eq. (12).

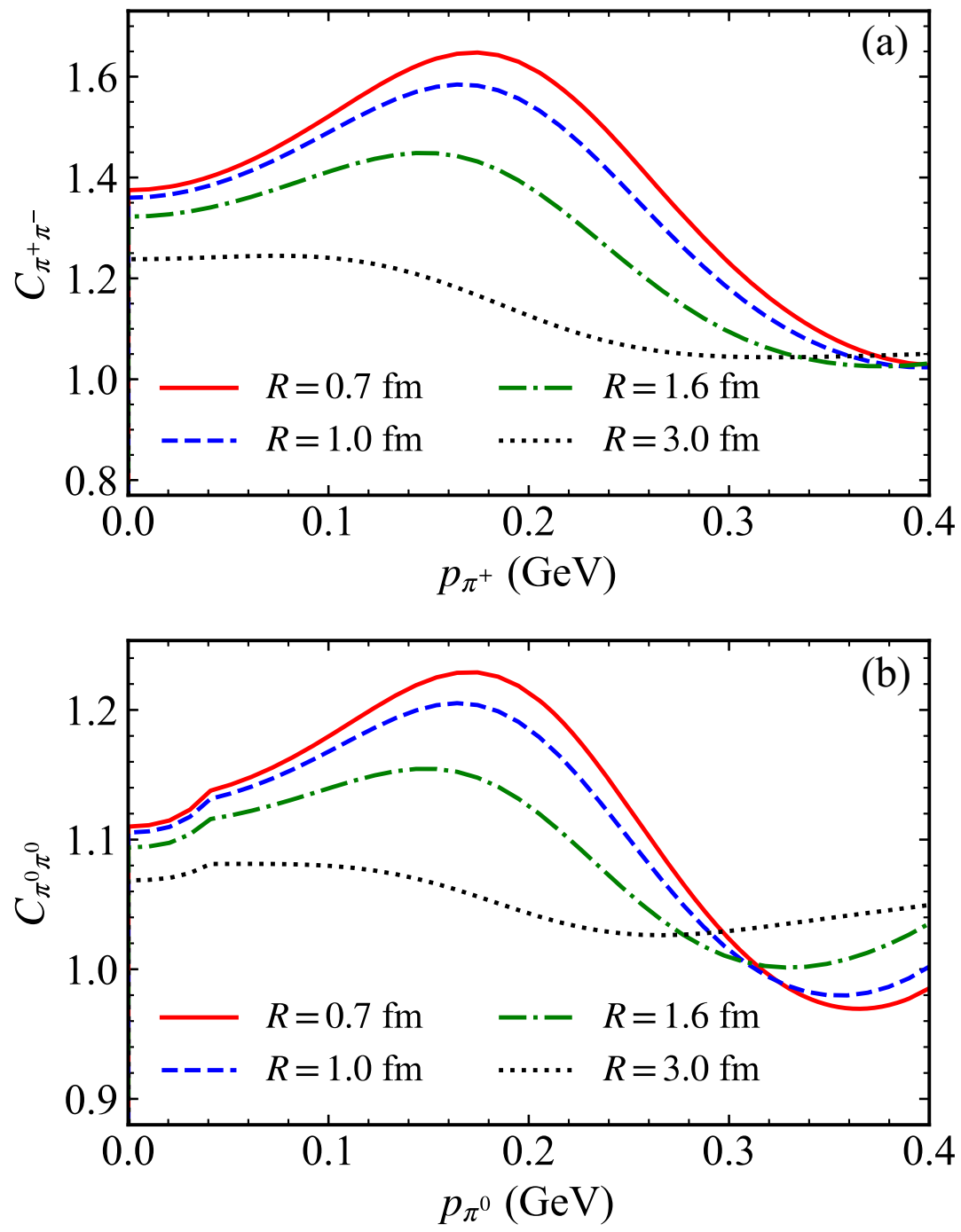

FIG. 6. Source size dependence of the correlation functions with the Gaussian source for $\pi^+\pi^-$ (a) and $\pi^0\pi^0$ (b) in coupled channels, taking $q_{\rm max} = 0.6$ GeV. The results obtained with source sizes $R = 0.7$, 1.0, 1.6, and 3.0 fm are denoted by the red solid line, the blue dashed line, the green dash-dotted line, and the black dotted line, respectively.

In Eqs. (16)–(17), the source size $R$ dependence of the coupled-channel correlation functions for $\pi^+\pi^-$ and $\pi^0\pi^0$ is shown in Figs. 6(a) and 6(b), respectively, as a function of the pion momentum. The overall behavior of both results is quite consistent, with each reaching its maximum when the pion momentum is around 0.18 GeV. For $\pi^+\pi^-$ and $\pi^0\pi^0$ pairs, the strength of correlations decreases gradually as the source size $R$ increases. A slight cusp structure is observed in the $\pi^0\pi^0$ correlation, resulting from the $\pi^+\pi^-$ threshold effect. As for the difference between the correlation functions of charged and neutral pions, in addition to the mass difference, the electromagnetic interaction between charged pions also contributes. However, this interaction was not accounted for in our calculation. The femtoscopy correlation functions in the presence of Coulomb potential have been studied in Ref. [56], which provided a more detailed discussion. These phenomena may be probed experimentally in future studies and could shed light on hadronic interactions.

In summary, the recent discovery of a substructure near the $\pi^+\pi^-$ mass threshold in $\psi(3686) \to J/\psi\pi^+\pi^-$ by the BESIII Collaboration challenges conventional interpretations of the dipion invariant mass spectrum. We show that this structure can be explained as a direct manifestation of dipion correlation. Using a reaction model based on flavor symmetry and a chiral unitary approach to treat dipion correlation, we successfully reproduce the observed feature, establishing strong evidence of dipion correlation in heavy quarkonium decays. The measured line shape is shown to stem from the interference between two distinct, yet equally important, OZI-suppressed decay mechanisms. Furthermore, the method enables us to calculate the dipion correlation function with these fixed parameters. This finding opens a new avenue for studying matter correlations and offers valuable insight into the nonperturbative dynamics of the strong interaction in the low-energy regime.

We would like to thank Professor De-Liang Yao for his valuable comments. This work is supported by the Natural Science Special Research Foundation of Guizhou University Grant No. 2024028 and Guizhou Provincial Major Scientific and Technological Program No. XKBF (2025)010 (Z. Y. W.). This work is also supported by the National Natural Science Foundation of China under Grants No. 12335001 and No. 12247101, the "111 Center" under Grant No. B20063, the Natural Science Foundation of Gansu Province (Grants No. 22JR5RA389, No. 25JRRA799), the Talent Scientific Fund of Lanzhou University, the fundamental Research Funds for the Central Universities (Grant No. lzujbky-2023-stlt01), the project for top-notch innovative talents of Gansu province, and Lanzhou City High-Level Talent Funding.

*Data availability.* The data are not publicly available. The data are available from the authors upon reasonable request.